\input psfig.sty

\def\ptitle{General energy bounds for systems of bosons with soft cores}
\nopagenumbers
\hsize 6.0 true in 
\hoffset 0.25 true in 
\emergencystretch=0.6 in                 
\vfuzz 0.4 in                            
\hfuzz  0.4 in                           
\vglue 0.1true in
\mathsurround=2pt                        
\topskip=20pt                            
\def\nl{\noindent}                       
\def\np{\hfil\vfil\break}                
\def\title#1{\bigskip\noindent\bf #1 ~ \trr\smallskip} 
\font\trr=cmr12                         
\font\bf=cmbx12                         
\font\sl=cmsl12                         
\font\it=cmti12                         
\font\trbig=cmbx12 scaled 1500          
\font\tiny=cmr10                         
\def\ma#1{\hbox{\vrule #1}}             
\def\mb#1{\hbox{\bf#1}}                 
\def\ng{>\kern -9pt|\kern 9pt}          
\def\bra{{\rm <}}                       
\def\ket{{\rm >}}                       
\def\hi#1#2{$#1$\kern -2pt-#2}          
\def\hy#1#2{#1-\kern -2pt$#2$}          

\def\half{{1 \over 2}}
\def\quar{{1 \over 4}}


\output={\shipout\vbox{\makeheadline
                                      \ifnum\the\pageno>1 {\hrule}  \fi 
                                      {\pagebody}   
                                      \makefootline}
                   \advancepageno}

\headline{\noindent {\ifnum\the\pageno>1 
                                   {\tiny \ptitle\hfil page~\the\pageno}\fi}}
\footline{}
\newcount\zz  \zz=0  
\newcount\q   
\newcount\qq    \qq=0  

\def\pref #1#2#3#4#5{\frenchspacing \global \advance \q by 1     
    \edef#1{\the\q}
       {\ifnum \zz=1 { %
         \item{[\the\q]} 
         {#2} {\bf #3},{ #4.}{~#5}\medskip} \fi}}

\def\bref #1#2#3#4#5{\frenchspacing \global \advance \q by 1     
    \edef#1{\the\q}
    {\ifnum \zz=1 { %
       \item{[\the\q]} 
       {#2}, {\it #3} {(#4).}{~#5}\medskip} \fi}}

\def\gref #1#2{\frenchspacing \global \advance \q by 1  
    \edef#1{\the\q}
    {\ifnum \zz=1 { %
       \item{[\the\q]} 
       {#2}\medskip} \fi}}

 \def\sref #1{~[#1]}

\def\references#1{\zz=#1
   \parskip=2pt plus 1pt   
   {\ifnum \zz=1 {\noindent \bf References \medskip} \fi} \q=\qq

 \pref{\thira}{W. Thirring, Found. Phys.}{20}{1103, (1990)}{}
 \bref{\thirb}{W. Thirring}{A Course in Mathematical Physics 4: Quantum Mechanics of Large Systems}{Springer, New York, 1983}{}
  \pref{\halla}{R. L. Hall, Phys. Rev. A}{45}{7682 (1992)}{}
  \pref{\hallaa}{R. L. Hall, Phys. Rev. A}{51}{3499 (1995)}{}
\pref{\fow}{R. H. Fowler and H. Jones, Proc. Camb. Phil. Soc.}{34}{573 (1938)}{}
\pref{\pen}{O. Penrose and L. Onsager, Phys. Rev.}{104}{576 (1956)}{}
\pref{\baym}{G. Bayme and C.J. Pethick, Phys. Rev. Lett.}{76}{6 (1996)}{}
\pref{\shur}{E.V.Shuryak, Phys. Rev. A}{54}{3151 (1996)}{}
\pref{\stoof}{H. T. C. Stoof, J. Stat. Phys.}{87}{1353 (1997)}{}
\pref{\perez}{V. M. P\'erez-Garcia, H. Michinel, J. I. Cirac, M. Lewenstein, and P. Zoller, Phys. Rev. A}{56}{1424 (1997)}{}
\pref{\fetter}{A. L. Fetter, J. Low Temp. Phys. A}{106}{643 (1997)}{}
\pref{\ueda}{M. Ueda and A. J. Leggett, Phys. Rev. Lett.}{80}{1576 (1998)}{}
\pref{\bohn}{J. I. Bohn, B.D.Esry, and C. H. Greene, Phys. Rev. A}{58}{584 (1998)}{}
\pref{\lieb}{E. H. Lieb and J. Yngvason, Phys. Rev. Lett.}{80}{2504 (1998)}{}
\pref{\lem}{L. F. Lemmens, F. Brosens, and J. T. Devreese, Phys. Rev. A}{59}{3112 (1999)}{}
\pref{\zal}{M. A. Zaluska-Kotur, M. Gajda, A. Orlowski, and J. Mostowski, Phys. Rev. A}{61}{033613 (2000)}{}
\pref{\temp}{J. Tempere, F. Brosens, L. F. Lemmens, and J. T. Devreese, Phys. Rev. A}{61}{043605 (2000)}{}
\pref{\cala}{F. Calogero, J. Math. Phys.}{10}{2191 (1969)}{}
\pref{\calb}{F. Calogero, J. Math. Phys.}{12}{419 (1971)}{}
\pref{\turb}{A. Turbiner, Mod. Phys. Lett. A}{13}{1473 (1998)}{}
\pref{\mcguire}{J. B. McGuire, J. Math. Phys.}{5}{622, (1964)}{}
\pref{\hallp}{R. L. Hall and H. R. Post, Proc. Phys. Soc (Lond.)}{90}{381 (1967)}{}
 \pref{\hallc}{R. L. Hall, Proc. Phys. Soc (Lond.)}{91}{16 (1967)}{}
\bref{\prug}{E. Prugovecki}{Quantum Mechanics in Hilbert Space}{Academic, New York, 1981}{}
  \bref{\reed}{M. Reed and B. Simon}{Methods of Modern Mathematical Physics IV: Analysis of Operators}{Academic, New York, 1978}{}
  \bref{\thirc }{W. Thirring}{A Course in Mathematical Physics 3: Quantum Mechanics of Atoms and Molecules}{Springer, New York, 1981}{}
\pref{\halle}{R. L. Hall, Can. J. Phys.}{50}{305 (1972)}{}
\pref{\hallf}{R. L. Hall, Aequ. Math.}{8}{281 (1972)}{}
\pref{\hallsa}{R. L. Hall and N. Saad, J. Phys. A}{32}{133 (1999)}{}
\pref{\kra}{A. Kratzer, Z. Physik}{3}{289 (1920)}{}
\pref{\fues}{E. Fues, Ann. Physik}{80}{367 (1926)}{}
\pref{\hooy}{G. van Hooydonk, J. Mol. Struct. (Theochem)}{109}{84 (1984)}{}
\pref{\sec}{D. Secrest, J. Chem. Phys.}{89}{1017 (1988)}{}
\pref{\req}{A. Requena, J. Zuniga, L. M. Fuentes and A. Hidolgo, J. Chem. Phys.}{85}{3939 (1986)}{}
\pref{\fra}{J. M. Frances, J. Zuifga, M. Alacid and A. Requena, J. Chem. Phys.}{90}{5536 (1989)}{}
\bref{\judd}{R. B. Judd}{Angular Momentum Theory for Diatomic Molecules}{New York: Academic Press,1975}{}
\pref{\sou}{A. de Souza Dutra, Phys. Rev. A}{47}{R2435 (1993)}{}
\pref{\nag}{N. Nag, R. Roychoudhury and Y. P. Varshni, Phys. Rev. A}{6}{5098 (1994)}{}
\pref{\znoj}{M. Znojil, Phys. Rev. A}{51}{128 (1995)}{}
\pref{\hallsb}{R. L. Hall and N. Saad, J. Chem. Phys.}{109}{2983 (1998)}{}
 
 }

\references{0}    

\trr 
\vskip 1.0true in
\centerline{\trbig General energy bounds for systems of bosons}
\vskip 0.3true in
\centerline{\trbig  with soft cores}
\vskip 0.5true in
\baselineskip 12 true pt 
\centerline{\bf Richard L. Hall}\medskip
\centerline{\sl Department of Mathematics and Statistics,}
\centerline{\sl Concordia University,}
\centerline{\sl 1455 de Maisonneuve Boulevard West,}
\centerline{\sl Montr\'eal, Qu\'ebec, Canada H3G 1M8.}
\vskip 0.2 true in
\centerline{email:\sl~~rhall@cicma.concordia.ca}
\bigskip\bigskip

\baselineskip = 17true pt  
\centerline{\bf Abstract}\medskip
We study a bound system of $N$ identical bosons interacting by model pair potentials of the form $V(r) = \lambda~{\rm sgn}(p)r^{p} + {{\mu}\over{r^2}},$  $\lambda > 0, \quad \mu\geq 0.$  By using a variational trial function and the `equivalent \hi{2}{body} method', we find explicit upper and lower bound formulas for the \hi{N}{particle} ground-state energy in arbitrary spatial dimensions $d \geq 3$ for the two cases $p = 2$ and $p = -1.$  It is demonstrated that the upper bound can be systematically improved with the aid of a special \hy{large}{N} limit in collective field theory.  
\medskip\noindent PACS~~03.65.Ge;~31.15.Pf;~03.75.Fi.
\np
    \title{1. Introduction}\smallskip
The principal subject of this paper is the ground-state energy of a system composed of $N$ identical bosons bound together by pair potentials.  Such systems usually collapse in the \hy{large}{N} limit; that is to say, the binding energy per particle rises with $N$ to infinity\sref{\thira-\thirb.}   In two earlier papers we studied gravitating boson systems\sref{\halla}, and systems of bosons or fermions which interact by wider classes of purely attractive pair potential\sref{\hallaa}.  In the present paper we extend this work for the boson case to systems with pair potentials which bind the system but have a soft repulsive core of the form ${\mu\over {r^2}}.$  The methods we develop to analyse these systems may have application to models for Bose-Einstein condensates\sref{\fow, \pen} in which there is at present much renewed interest\sref{\baym-\temp}.  As in our earlier work, the goal of the present study is to provide simple general energy bounds as functions of the parameters $m,$ $\lambda,$ $\mu,$ and $N.$  Since is not difficult to do so, we also allow for an arbitrary number $d\geq 3$ of spatial dimensions.\medskip

Formulas for \hi{N}{particle} upper and lower energy bounds are derived for pair potentials $V_{ij} = V_{o}f(r_{ij})$ whose shapes $f(r)$ are either harmonic oscillators with a soft core  $f(r) = \lambda r^2 + {\mu\over {r^2}},$ or Kratzer potentials $f(r) = -{{\lambda}\over{r}} + {\mu\over {r^2}}.$  The shapes of these two potentials are shown in Fig.(1) for $\lambda = \mu = 1.$  The first of the models could perhaps be considered as a generalization of the soluble \hi{1}{dimensional} Calogero model\sref{\cala - \turb} to dimension $d\geq 3.$   For such potentials, with minima $f(\hat{r})$ at $r = \hat{r} > 0,$ provided there are enough spatial dimensions available ($d+1\geq N$) for every pair distance to satisfy $r_{ij} = \hat{r},$ then, in classical mechanics the lowest energy would be equal to ${{N}\choose {2}}V_{o}f(\hat{r}).$  This expression provides a lower energy bound in both classical and quantum mechanics; our general energy bounds (valid for all $N\geq 2$) will show that this value is approached asymptotically for large $V_{o}$ in the limit $d\rightarrow\infty;$ this is possible because the positive contribution from zero-point oscillations varies like $(N-1)(NV_{o})^{\half}$ and is eventually dominated by the static potential term.\medskip

The non-individuality of identical quantum-mechanical particles introduces a very powerful constraint which allows us to relate the \hi{N}{body} problem to a specially constructed (reduced) \hi{2}{body} problem having an overall factor of $N-1$ and a potential enhanced by $N/2,$ corresponding respectively to the $N-1$ relative kinetic-energy operators and the $N(N-1)/2$ pair potentials.   This `equivalent two-body' notion, which is central to our approach to the study of these systems, will be formulated explicitly in Section~(2). The \hi{N}{body} energy ${\cal E}$ is described by function $E = F_{N}(v) = {\cal E}/(N-1),$ where $v = {N\over 2}V_o,$ and $V_o$ is the coupling parameter (in units $\hbar = m = 1$).  Since $v$ includes the factor $N,$ it follows that if we consider a {\it finite} value of $v$ and let $N\rightarrow\infty,$ this implies that $V_o\rightarrow 0.$  The well-known exactly-soluble \hi{\delta}{potential} problem in one dimension\sref{\mcguire} provides a convenient illustration of the energy function $F(v).$  If $V(r) = -V_o\delta(r),$ we have exactly
$${{\cal E}\over{N-1}} = E = F_{N}(v) = -{1\over 6}\left[1 + {1\over N}\right]v^2, \quad N \geq 2,\quad v = {N\over 2}V_o.\eqno{(1.1)}$$
The shape of the potential dictates the form $v^2$ of these energy curves, and there is a distinct curve for each value of $N\geq 2.$ The curves of this particular example satisfy the general functional inequalities\sref{\halla}
$$F_2(v) \leq E = F_N(v) \leq F_\infty(v) \leq F_{\phi}(v) \leq F_G(v),\eqno{(1.2)}$$
where, as will be explained later, $F_{\phi}(v)$ is an upper bound obtained with the aid of collective-field theory in the \hy{large}{N} limit, and $F_G(v)$ is the upper-bound curve obtained by employing an \hi{N}{body} translation-invariant Gaussian trial wave function. The ordering of the $F_N(v)$ curves with $N$ is a consequence of the monotonic increase in the severity of the boson-symmetrization constraint with increasing $N.$  For the soluble \hi{\delta}{potential} model we have\sref{\halla} explicitly: $F_\infty(1) = -1/6,$ $F_{\phi}(1) = -0.164868,$ and $F_G(1) = -1/2\pi.$ The family of densities used for the variational collective-field upper bound is given by
$$\phi(r) = e^{-(r/b)^{q}},\quad b,\ q > 0.\eqno{(1.3)}$$
\nl The energy is minimized with respect to the positive scale and power parameters $b$ and $q.$ The effectiveness of this useful \hi{2}{parameter} family of trial densities is also demonstrated by the applications discussed in Sections (4) and (5) below.   We conjecture that $F_{\phi}(v)$ is close to $F_{\infty}(v)$ for the more general problems considered in this paper but we know no way of proving such a claim at this time. For another well-known soluble problem, the \hi{d}{dimensional} harmonic oscillator $V(r) = V_{o}r^2,$ with , $d\geq 1,$  the inequalities (1.2) all collapse together to the common exact value $E = d v^{\half}.$    
    \title{2. The \hi{N}{body} problem in the centre-of-mass frame}\smallskip
The Hamiltonian, with center-of-mass removed, for a system of $N$ identical particles each of mass $m$ interacting via central pair potentials may be written
$${\cal H} ={1 \over {2m}}\sum_{i = 1}^{N}\mb{p}_{i}^{2} -
{1 \over {2mN}}\left(\sum_{i = 1}^{N}\mb{p}_{i}\right)^{2} +
  \sum_{j>i=1}^{N}V_{o}f\left({{|\mb{r}_{i} -\mb{r}_{j}|} \over a}\right),\eqno{(2.1)}$$
where $V_{o}$ and $a$ are respectively the depth and range parameters of the potential with shape $f.$  By algebraic rearrangement (2.1) may be rewritten in the more symmetrical form 
$${\cal H} =\sum_{j>i=1}^{N}\left\{{1 \over {2mN}}(\mb{p}_{i} - \mb{p}_{j})^{2} + V_{o}f\left({{|\mb{r}_{i} -\mb{r}_{j}|} \over a}\right)\right\},\eqno{(2.2)}$$
We now define new coordinates by $\rho = BR,$ where $\rho = [\rho_{i}]$ and $R = [\mb{r}_{i}]$  are column vectors of the new and old coordinates, respectively, and $B$ is a real constant $N{\rm x}N$ matrix.  For convenience we require all the rows of $B$ to be unit vectors, we let the elements of the first row all be equal to ${1 \over \sqrt{N}},$ so that $\rho_{1}$ is proportional to the centre-of-mass coordinate; we also require that the remaining $N-1$ rows of $B$ be orthogonal to the first row, so that they define a set of $N-1$ relative coordinates.  One more row is also fixed so that we have at least one pair distance at our disposal, namely
$$\rho_{2}= {{\mb{r}_{1} - \mb{r}_{2}} \over \sqrt{2}}.\eqno{(2.3)}$$
For boson systems, we have found that Jacobi relative coordinates, for which $B$ is orthogonal, are the most useful.  Thus, corresponding to the transformation $\rho = BR$ of the coordinates, it follows that the column vector $P$ of the associated momenta transforms to the new momenta $\Pi = [\pi_{i}]$ by the relation $\Pi = (B^{T})^{-1}P = P.$  If $\Psi$ is any translation-invariant wave function for the \hi{N}{body} system composed of identical bosons, then we can write\sref{\hallp, \hallc} the following mean energy relation between the \hi{N}{body}  and \hi{2}{body} systems:
$$(\Psi, {\cal H}\Psi) = (\Psi, \ma{H} \Psi),\eqno{(2.4)}$$
where the `reduced'  two-particle Hamiltonian $\ma{H}$ is given by
$$\ma{H} = (N-1)\left({1 \over {2m}}\pi_{2}^{2} +
    {N \over 2}V_{o}f\left({{\sqrt{2}|\rho_{2}|} \over a}\right)\right).\eqno{(2.5)}$$

Further simplifications can be achieved if we work with dimensionless quantities.  We suppose that the translation-invariant \hi{N}{body} energy is ${\cal E}$ and we define the dimensionless energy and coupling parameters $E$ and $v$ by the expressions
$$E = {{m{\cal E}a^{2}} \over {(N-1)\hbar^{2}}},\quad
            v = {{NmV_{o}a^{2}} \over {2\hbar^{2}}}.\eqno{(2.6)}$$
It is then natural to define a dimensionless versions of the reduced \hi{2}{body} Hamiltonian $\ma{H}$ and the relative coordinate $\rho_{2}$ by the relations
$$H ={{m\ \ma{H}a^{2}} \over {(N-1)\hbar^{2}}}= -\Delta + vf(r),\quad 
\mb{r} = \sqrt{2}\rho_{2}/a = (\mb{r}_{1} - \mb{r}_{2})/a,\quad r = \|\mb{r}\|.\eqno{(2.7)}$$
We note that the Hamiltonian $H$ depends on $N$ {\it only} through the dimensionless coupling parameter $v.$  By the Rayleigh-Ritz (min-max) principle\sref{\prug-\thirc}, we have the following characterization of the \hi{N}{body} ground-state energy parameter $E$ in terms of $H:$
$$E = \min_{\Psi}{{(\Psi, H\Psi)} \over {(\Psi, \Psi)}} = F_{N}(v),\eqno{(2.8)}$$
where $\Psi$ is a translation-invariant function of the $N-1$ relative coordinates (and spin variables, if any) which is symmetric under the permutation of the $N$ individual-particle indices.  The \hi{N}{body} energy ${\cal E}$ is recovered from $E$ by inverting (2.6). Thus we have explicitly:
$${\cal E} = {{(N-1)\hbar^{2}} \over {ma^{2}}}F_{N}\left({{NmV_{o}a^{2}} \over {2\hbar^{2}}}\right).\eqno{(2.9)}$$
However, for the remainder of this paper we shall work with the dimensionless form $E = F_{N}(v):$ the problem is to find or approximate $F_{N}.$   In each situation we shall have to specify the dimension $d$ of physical space; usually $d = 3.$.

 \title{3. Energy bounds}\smallskip
The energy bounds used in this paper are summarized in terms of the $F$ functions by Eq.(1.2). The history of the equivalent \hi{2}{body} method for boson systems has been described in the earlier papers\sref{\halla, \hallaa} and in the references therein.  The main result is a general energy lower bound which, for boson systems with orthogonal Jacobi relative coordinates, is given by 
$$F_{2}(v)\leq E  = F_{N}(v),\eqno{(3.1)}$$
where $F_{2}(v)$ is the lowest eigenvalue of the \hi{1}{particle} (`reduced' \hi{2}{particle}) Hamiltonian $H = -\Delta + vf(r).$  With equal simplicity, our weakest upper bound $F_{G}(N),$  provided by a  Gaussian boson trial function $\Psi,$ may also be expressed in terms of the \hi{1}{body} operator $H.$  This is a consequence of the following argument. If and only if the symmetric translation-invariant function $\Psi$ is Gaussian\sref{\halle, \hallf}, it may be factored in the form
$$\Psi(\rho_2,\rho_3,\dots,\rho_N) = \psi(\rho_2)\eta(\rho_3,\rho_4, \dots, \rho_N).\eqno{(3.2)}$$
But the equivalence (2.4) then implies, in this case, that $E \leq (\psi,H\psi)||\psi||^{-2}.$  This explains why the inequalities (1.2) collapse together in the case of the harmonic oscillator for which the exact \hi{1}{body} lowest eigenfunction of $H$ is also Gaussian. In this argument we assume for the upper bound that $\bra H \ket$ has been optimized with respect to the scale of the wave function. The boson symmetry of these Gaussian functions is demonstrated most clearly by the following algebraic identity:
$$\sum_{j > i = 1}^{N}(\mb{r}_{i} - \mb{r}_{j})^{2} = N \sum_{k = 2}^{N}\rho_{k}^{2}.\eqno{(3.3)}$$ 

The potentially better upper bound $F_{\phi}(N)$ is more complicated both to derive and to compute\sref{\halla}. It is found by considering the collective field model in the \hy{large}{N} limit.  The energy so obtained is an upper bound to $F_{\infty}(v)$ and this, in turn, may be closely approximated from above by optimizing the right-hand side of the following equation:
$$F_{N}(v) \leq F_{\phi}(v) = {1\over 8}\int{{(\nabla\phi(r))^{2}}\over{\phi(r)}}d^{d}r + v \int\int\phi(r)f(|r - r'|)\phi(r')d^{d}r d^{d}r',\eqno{(3.4)}$$
where $\phi(r)$ is a trial probability density function (for inter-particle distances) satisfying $\int\phi(r)d^{d}r = 1.$ It has been shown\sref{\halla} that if $\phi$ is Gaussian, and the energy is optimized with respect to a scale parameter, then the result $F_{\phi}(v)$ is identical to the upper bound $F_{G}(v)$ obtained with the aid of a scale-optimized translation-invariant Gaussian trial wave function.  The wider possible choice of the form of the probability density $\phi$ allows us to transcend the Gaussian bound whilst still working essentially with a variational `function' of single variable for the \hi{N}{body} problem.

In order to facilitate the reproduction of our results and the application of the method to other problems, we make the following explicit technical remarks concerning the case $d = 3.$   It is very helpful to think of the probability density as a  function $\phi(r) = w(r/b) = w(s),$ which depends on the remaining parameter $q,$ to be discussed later.  If we let $I = \int_{0}^{\infty}w(s)s^2ds,$ then the kinetic energy integral becomes 
$$\bra KE \ket = {1\over{8Ib^2}}\int_0^{\infty}{{(w'(s))^{2}}\over{w(s)}}s^2ds\eqno{(3.5)}$$
\nl and potential energy integrals for pure powers may be written
$$\bra r^p\ket = {{b^{p}}\over{I^{2}(p+2)}}\int_{0}^{\infty}dt\ w(t)t\int_{t}^{\infty}ds\  w(s)s\left\{(s+t)^{p+2} - (s-t)^{p+2}\right\},\quad p \neq -2.\eqno{(3.6)}$$
For the case $p = -2$ we have instead
$$\bra {1\over{r^2}}\ket = {1\over{I^{2}b^{2}}}\int_{0}^{\infty}dt\ w(t)t\int_{t}^{\infty}ds w(s)s\ln\left({{s+t}\over{s-t}}\right).\eqno{(3.7)}$$
\nl With the terms expressed in this form, the minimization over scale $b > 0$ can often be carried out explicitly yielding an algebraic expression which then needs to be minimized with respect to the remaining parameter $q.$  We have explored various alternative forms for $w(s),$ such as $s^q \exp(-s^2)$ and $\exp(-(s-q)^2),$ but have found it difficult to improve on the variable-power family $w(s) = \exp(-s^q)$ which we have used to obtain the results discussed in detail in Sections (4) and (5) below. 
 
\title{\bf 4. The harmonic oscillator with a soft core}\smallskip
We now choose the potential shape to be
$$f(r) = \lambda r^2 + {\mu\over {r^2}},\quad \lambda > 0,\quad \mu\geq 0.\eqno{(4.1)}$$
The lower bound $F_{2}(v)$ is given by the lowest eigenvalue of $H = -\Delta + vf(r)$ in $d$ dimensions, a problem which is discussed, for example, in Ref.\sref{\hallsa}.  The Gaussian upper bound $F_{G}(v)$ is provided by minimizing the Rayleigh quotient $(\psi, H\psi)/(\psi, \psi)$ with respect to the scale variable $\alpha$ in $\psi(r) = e^{-\half\alpha r^{2}}$ in $d$ dimensions. These calculations yield the following bounds on the \hi{N}{boson} energy parameter $E$ (the energy  itself is recovered essentially by multiplying $E$ by $N-1,$ according to Eq.(2.10)) 
$$2(v\lambda)^{\half}\left[1 + \left(\mu v + ({d\over 2}-1)^{2}\right)^{\half}\right]\ <\ E \quad<\quad (d v\lambda)^{\half}\left[d + {{4v\mu}\over{d-2}}\right]^{\half},\quad d \geq 3.\eqno{(4.2)}$$
\nl The Gaussian trial wave function allows us to compute an approximate value for the mean-squared pair separation $\sigma^{2},$ a measure of the size of the system. We find
$$\sigma^{2} = \bra (\mb{r}_1 - \mb{r}_2)^{2}\ket = {{d}\over{2(v\lambda)^{\half}}}\left(1 + {{4v\mu}\over{d(d-2)}}\right)^{\half}.\eqno{(4.3)}$$

We now look at some special cases.  For the harmonic oscillator, $\mu = 0,$ the inequalities collapse to the exact value $E = d(v\lambda)^{\half}.$   The asymptotic forms of the bounds as $\Rightarrow v\rightarrow\infty$ and $\mu > 0$ are given by

$$\sim 2v(\lambda\mu)^{\half}\  <\  E\  <\  \sim 2v(\lambda\mu)^{\half}\left({d\over{d-2}}\right)^{\half}.\eqno{(4.4)}$$

\nl Thus for large $N$ the energy per particle $E$ increases like $N,$ although the `size' of the system (as estimated by the Gaussian wave function) approaches the constant value $\sigma = (d/(d-2))^{\quar}(\mu/\lambda)^{\quar}.$  If $d$ is now taken large, the asymptotic form of the energy approaches the classical expression $E = vf(\hat{r}),$ where $\hat{r} = \sigma = (\mu/\lambda)^{\quar}$ is the position of the minimum of $f(r).$

For the special case $d = 3$ we obtain in general
$$(v\lambda)^{\half}\left[2 + \left(1 + 4\mu v\right)^{\half}\right]\  < \  E\  < \  (v\lambda)^{\half}\left[9 + 12v\mu\right]^{\half}.\eqno{(4.5)}$$
\nl These results are shown in Fig.(2) along with the improved upper bound (`dashed' curve) obtained with the trial probability density function (1.3).  The extreme optimal $q(v)$ values in the range shown were: $q(2) = 2.8593,\ q(20) = 4.460.$  Hence the optimal probability density $\phi(r)$ is found to be quite far from Gaussian.

\title{\bf 5. The Kratzer potential}\smallskip
The Kratzer potential\ has shape function given\sref{\kra-\znoj} by
$$f(r) = -{{\lambda}\over r} + {\mu\over {r^2}},\quad \lambda > 0,\quad \mu\geq 0.\eqno{(5.1)}$$
The \hi{N}{particle} lower bound $F_{2}(v)$ is provided by the lowest eigenvalue of $H = -\Delta + vf(r)$ in $d$ dimensions, a problem which is discussed, for example, in Ref.\sref{\hallsb}. Meanwhile an exactly similar calculation to that described in Section (4), with a Gaussian trial function, generates the corresponding upper bound.  We find in this way that the energy parameter $E$ satisfies the following inequalities:
$$-{{(v\lambda)^2}\over{\left[1+\left((d-2)^2 + 4v\mu\right)^{\half}\right]^{2}}}\  <\  E\  <\  -{{(v\lambda\gamma_{d})^2}\over{\left[2d + {{8v\mu}\over{d-2}}\right]}}~,\quad d\geq 3,\eqno{(5.2)}$$
where the \hi{d}{dependent} constant $\gamma_{d}$ is given by
$$\gamma_{d} = {{\Gamma\left({{d-1}\over 2}\right)}\over{\Gamma\left({d\over 2}\right)}},\quad\quad \gamma_{3} = {2\over\sqrt \pi}.\eqno{(5.3)}$$
\nl The Gaussian trial wave function again allows us to compute an approximate value for the mean-squared pair separation.  We find 
$$\sigma^2 = \bra (\mb{r}_1 - \mb{r}_2)^{2}\ket = {{d^{3}}\over{2(v\lambda\gamma_{d})^{2}}}\left(1 + {{4v\mu}\over{d(d-2)}}\right)^{2}.\eqno{(5.4)}$$

For the pure gravitational case $\mu = 0$ the energy inequalities become
$$-{{(v\lambda)^2}\over{(d-1)^2}} < E < -{{(v\lambda\gamma_{d})^2}\over{2d}}.\eqno{(5.5)}$$

\nl The asymptotic forms of the bounds as $\Rightarrow v\rightarrow\infty$ and $\mu > 0$ are given by
$$\sim - {{v\lambda^{2}}\over{4\mu}}\  <\  E\  <\  \sim - {{v\lambda^{2}}\over{4\mu}}M(d),\quad {\rm where}\  M(d) = \left[{{\gamma_{d}^{2}(d-2)}\over 2}\right].\eqno{(5.6)}$$

\nl The function $M(d)$ increases monotonically with $d$ to 1; $M(3) = 2/\pi \approx 0.63662;$ $M(8) > 0.9.$  Thus, as for the previous model, the energy per particle $E$ increases like $N.$  Meanwhile the size (as estimated by the Gaussian wave function) approaches the constant value $\sigma = (2\sqrt{2}d/(\gamma_{d}(d-2))(\mu/\lambda).$ If $d$ is now taken large, the asymptotic form of the energy (again) approaches the classical expression $E = vf(\hat{r}),$ in which $\hat{r} = \sigma = 2\mu/\lambda$ is the position of the minimum of $f(r).$

For the special case $d = 3$ we obtain in general
$$-{{(v\lambda)^2}\over{\left[1+\left(1 + 4v\mu\right)^{\half}\right]^{2}}}\  <\  E\  <\  -{{(v\lambda)^{2}}\over{\pi\left[{3\over 2} + 2v\mu\right]}}.\eqno{(5.7)}$$
\nl The bounds for three dimensions are shown in Fig.(3), along with the improved
 upper bound (`dashed' curve) obtained with the trial probability density function (1.3).  The extreme optimal $q(v)$ values in the range shown were for this problem: $q(2) = 2.0017,\ q(20) = 3.237.$  The optimal probability density (in this family) is found to be almost Gaussian for small $v < 3$ but very different from Gaussian for larger values of $v.$

\title{\bf 6. Conclusion}\smallskip
The main purpose of this paper is to derive general bounds for the energy of 
\hi{N}{boson} systems which are bound together by pair potentials with soft cores.
  We have examined two such models and we have provided upper and lower bound formulas
 valid for all $N\geq 2$ and $d\geq 3,$ and for all values of the potential parameters
 that bind the system. These bounds are expressed in terms of the energy parameter
 $E = {\cal E}/(N-1) = F_{N}(v),$ where (in units with $\hbar = m = 1$) $v = NV_o/2.$
 If the potential shape is such that $F_{2}(v)$ is close to $F_{\infty}(v),$ then we
 obtain close upper and lower bounds valid for all $N\geq 2.$  The  upper bound
 $F_{G}(v)$ provided by a Gaussian trial function may be improved to $F_{\phi}(v)$
 which is derived by using a (possibly non-Gaussian) trial probability density
 $\phi$ in a limiting form of collective field theory. In order to do significantly
 better one might hand craft a trial wave function for a particular $f(r)$ and $N.$ 
 Of course, to be secure about this wave function one would still need to find a
 good lower bound, a goal usually not easy to achieve.  The global results we have
 obtained show that these boson systems are asymptotically bounded by expressions
 of the form ${\cal E} \sim cN^{2};$ if $d$ is also large, we have shown that the
 heuristic classical expression ${\cal E} = {N\choose 2}V_{o}f(\hat{r})$ is reached
 asymptotically. Since the many-body problem continues to offer a serious challenge
 for direct numerical solution, it is very helpful to have some explicit analytic
 upper and lower energy bounds and size estimates for model systems such
 those discussed in this paper.     

   \title{Acknowledgment}
Partial financial support of this work under Grant No. GP3438 from the Natural Sciences and Engineering Research Council of Canada is gratefully acknowledged. 
\np
 \references{1}
\np
\baselineskip 18 true pt 

\hbox{\vbox{\psfig{figure=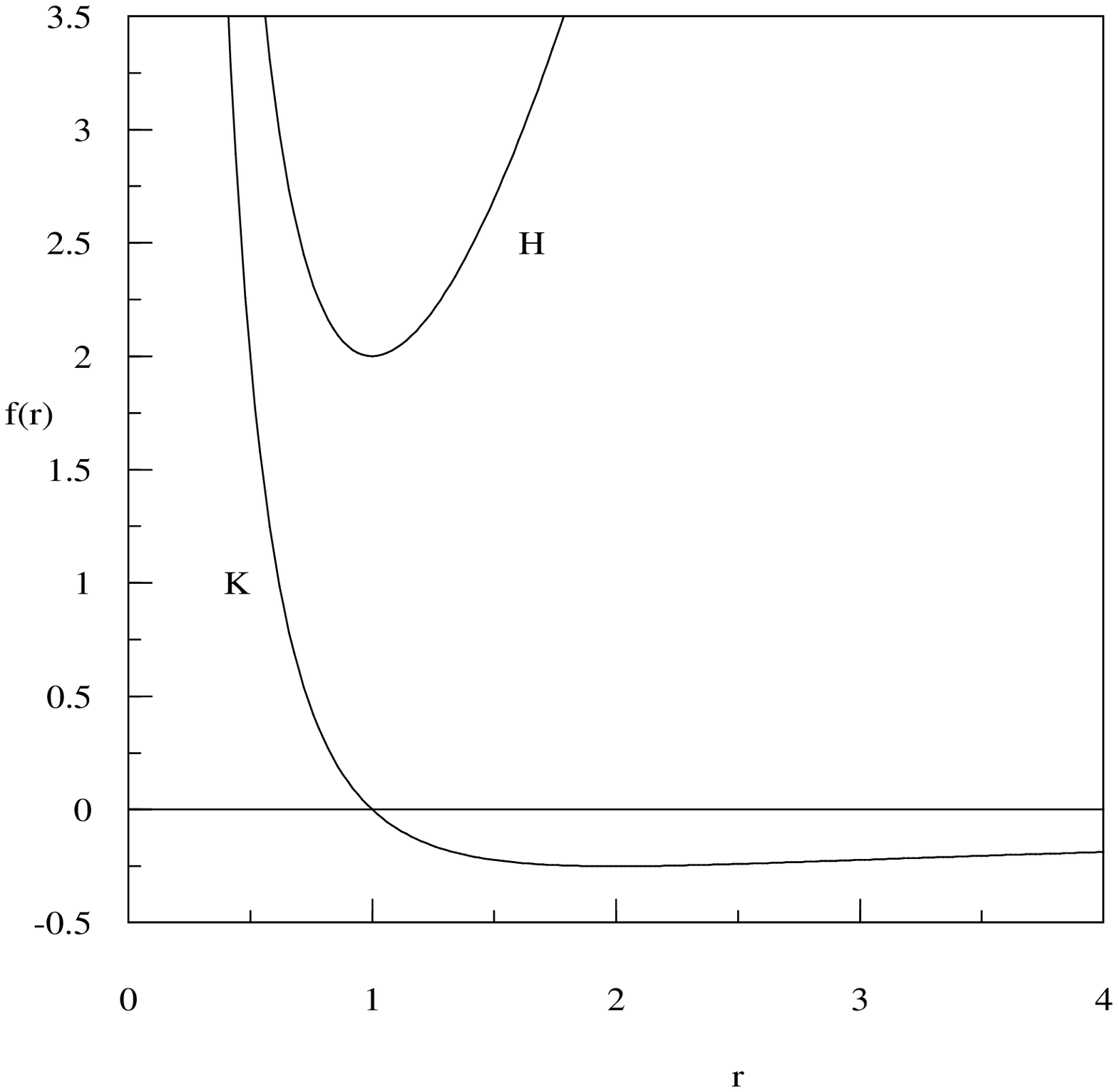,height=6in,width=5in,silent=}}}
\title{Figure 1.}
\nl The shapes of the two potentials studied: the harmonic oscillator with soft core, $H = f(r) = \lambda r^{2} + \mu r^{-2},$ and Kratzer's potential, $K = f(r) = -\lambda r^{-1} + \mu r^{-2},$ with $\lambda = \mu = 1.$
\np
\hbox{\vbox{\psfig{figure=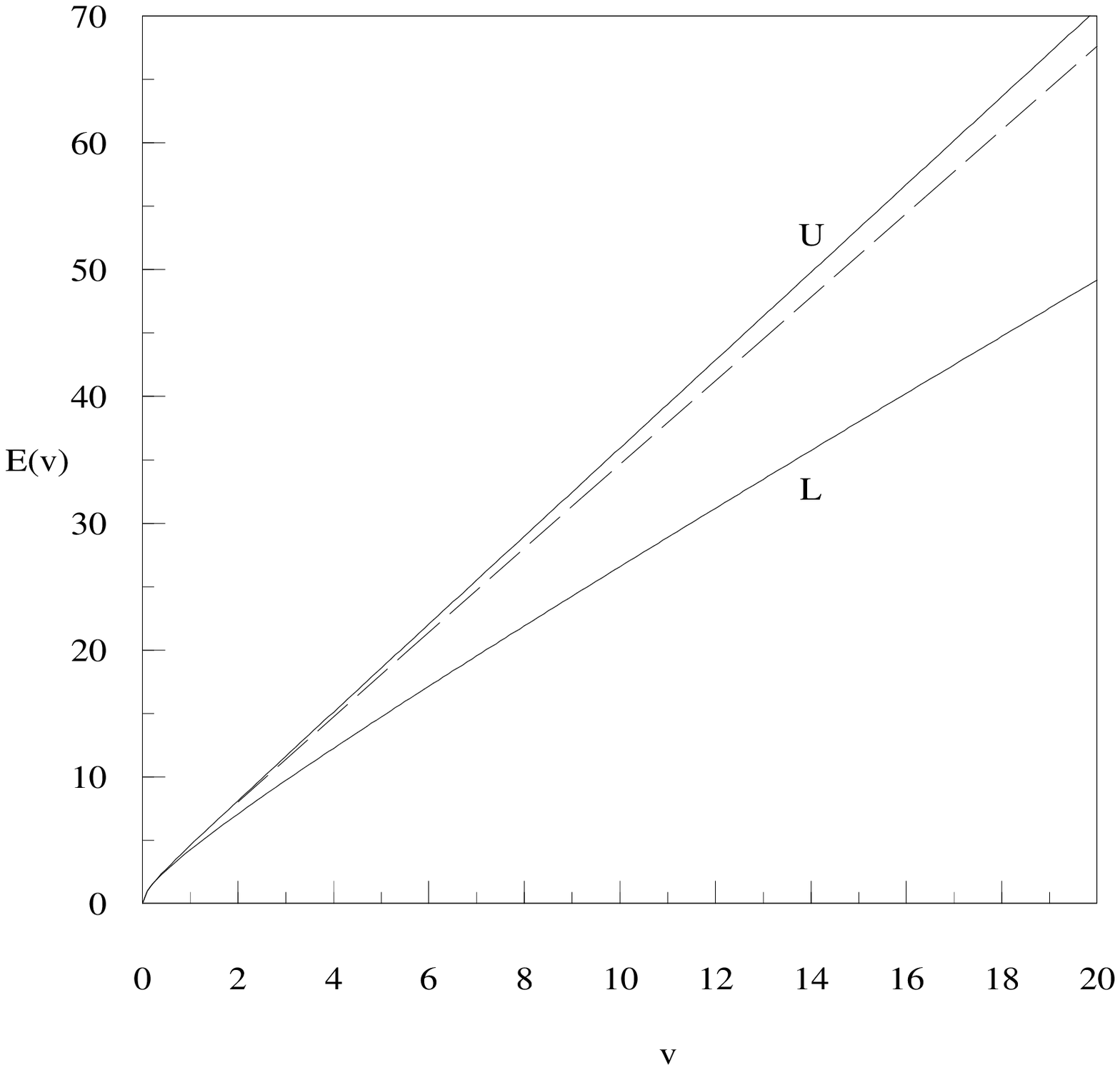,height=6in,width=5in,silent=}}}

\title{Figure 2.}
\nl Bounds for the energy parameter $E(v) = {\cal E}/(N-1)$ of the \hi{N}{boson} system with pair potentials of the form $f(r) = V_o(r^{2} +  r^{-2}),$ as a function of $v = NV_0/2.$ The graphs show: U the upper bound $F_{G}(v)$ found by a Gaussian trial function, L the lower bound $F_{2}(v)$ by the equivalent \hi{2}{body} method, and `dashed' the upper bound $F_{\phi}(v)$ by collective field theory.  
\np
\hbox{\vbox{\psfig{figure=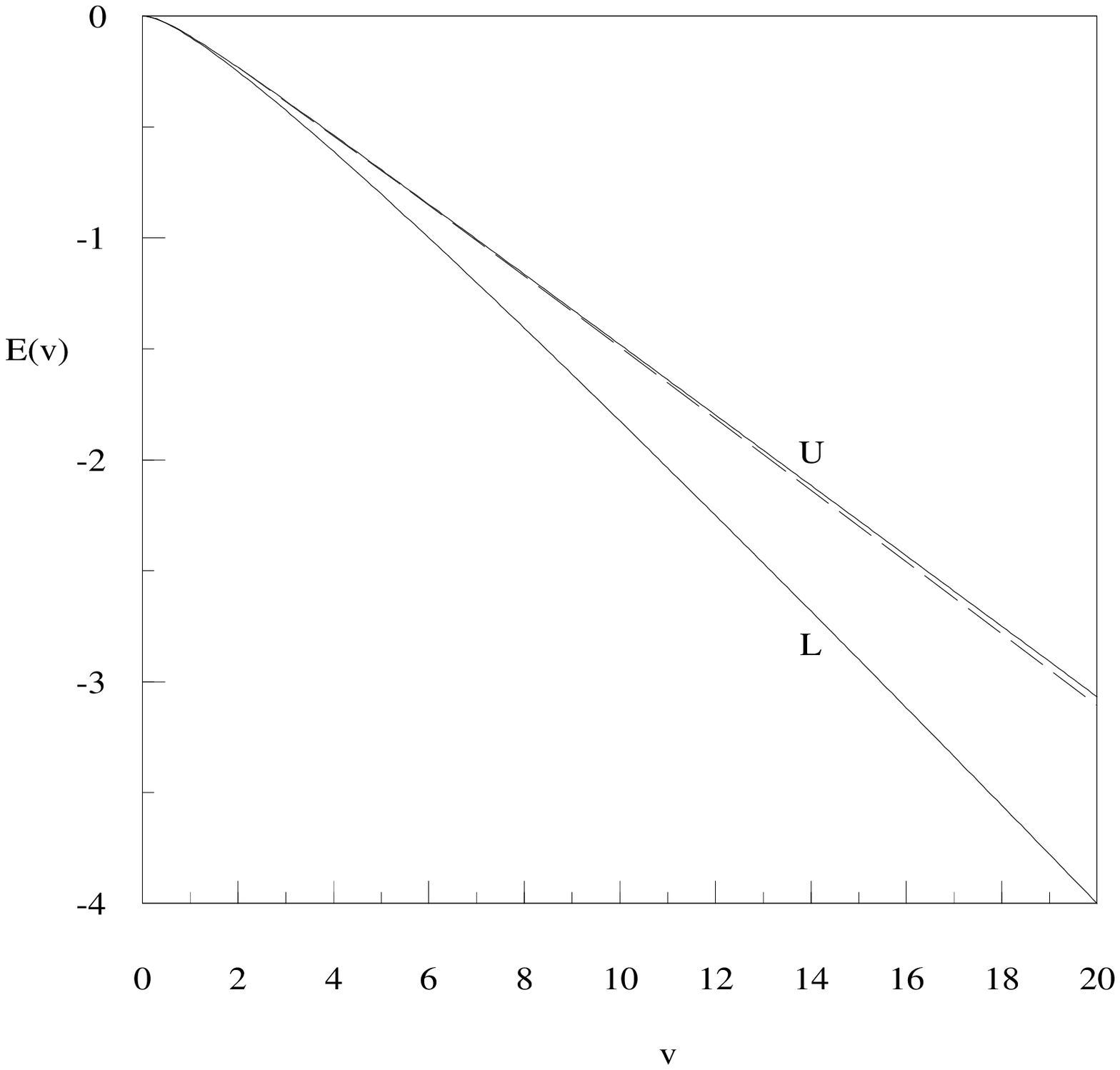,height=6in,width=5in,silent=}}}

\title{Figure 3.}
\nl Bounds for the energy parameter $E(v) = {\cal E}/(N-1)$ of the \hi{N}{boson} system with pair potentials of the form $f(r) = V_o( -r^{-1} + r^{-2}),$ as a function of $v = NV_0/2.$ The graphs show: U the upper bound $F_{G}(v)$ found by a Gaussian trial function, L the lower bound $F_{2}(v)$ by the equivalent \hi{2}{body} method, and `dashed' the upper bound $F_{\phi}(v)$ by collective field theory.

\end